\documentclass[aps,prb,twocolumn]{revtex4}
\usepackage{amsmath, amssymb, graphics, setspace}
\usepackage{amsmath}
\usepackage{graphicx}
\usepackage{dsfont}
\usepackage{multirow}

\begin{document}

\title{Spin-valley antiferromagnetism and topological superconductivity in
the trilayer graphene Moire super-lattice}
\author{Guo-Yi Zhu$^{1}$, Tao Xiang$^{2,3}$, and Guang-Ming Zhang$^{1,3}$}
\affiliation{$^{1}$State Key Laboratory of Low-Dimensional Quantum Physics and Department
of Physics, Tsinghua University, Beijing 100084, China \\
$^{2}$Institute of Physics, Chinese Academy of Sciences, Beijing 100190,
China \\
$^{3}$Collaborative Innovation Center of Quantum Matter, Beijing, China}
\date{\today}

\begin{abstract}
A recent experiment has shown that exotic correlated insulating phases
emerge in the ABC-stacked trilayer graphene-boron nitride Moire
super-lattice at both quarter and half-filling. A single-band minimal model
with valley contrasting staggered-flux is proposed to capture the relevant
band structure of this system, where the conspiracy of perfect Fermi-surface
nesting and van Hove singularity strongly enhance the valley fluctuation,
leading to inter-valley spiral (IVS) order at half filling. Nevertheless,
the weak coupling theory is insufficient to account for the correlated
insulating state near quarter filling. In this paper, we consider a strong
coupling U(1)$_{v}\times $SU(2)$_{s}$ symmetric spin-valley model to obtain
the correlated insulating state and the pairing instability near quarter
filling. A significant ingredient in the strong coupling model is the
Dzyaloshinsky-Moriya like interaction inherited from the flux, which breaks
not only the valley SU(2)$_{v}$ symmetry but also the sub-valley spatial
reflection symmetry. We discuss all the possible long-range orders
stabilized by the effective spin-valley-exchange interactions, and it turns
out that the flux remarkably enhance the ferro-spin inter-valley 120$^{\circ
}$ order, which shares the same valley feature as the IVS order. Upon
doping, the leading pairing instability lies in the inter-valley channel
with a trigonally warped $p\pm ip$-wave form factor in the presence of the
sub-valley reflection symmetry breaking. Depending on the sign of Hund's
coupling, the total pairing state could be either spin singlet or triplet.
While the spin singlet chiral topological pairing state $(p\pm ip)_{\uparrow
\downarrow }-(p\pm ip)_{\downarrow \uparrow }$ is necessarily chiral, the
spin triplet topological pairing state could be chiral $(p\pm ip)_{\uparrow
\uparrow }+(p\pm ip)_{\downarrow \downarrow }$, or helical $(p\pm
ip)_{\uparrow \uparrow }+(p\mp ip)_{\downarrow \downarrow }$.
\end{abstract}

\maketitle

\section{Introduction}

Valley is a novel low energy degree of freedom commonly studied in
graphene-based systems, which may be viewed as an isospin \cite%
{QianNiu,Beenakker}. It carries rather rich topological consequences and
been actively explored \cite{HaldaneModel, KaneMele,
Morpurgo,Mele13valleyChern,Rappe,MacDonald10ABC,Martin,KoshinoMcCann}.
However, the valley degree of freedom rarely sets foot in the Mott physics
and high Tc superconductivity in the past. The Moire heterostructure gives a
new opportunity. In these heterostructures, the original lattice periodicity
is broken, and a Moire super-lattice emerges on a larger scale, which
efficiently suppresses the kinetic energy scale by folding the bands. As a
result, the valley degree of freedom that is highly nonlocal in original
lattice is now tamed as a local orbital in the Moire superlattice.
Meanwhile, the local interactions that were otherwise weak could possibly
come to dominate the kinetics and lead to correlated physics.

Very recently, the experiment of magic-angle twisted bilayer graphene (TBG)
successfully demonstrate this scenario. Twisting the graphene bilayer is one
efficient way to produce triangular Moire superlattice\cite%
{Yankowitz,Dean,Hunt,Ponomarenko,YangZhang,ShiWang}. Under the tiny magic
angle like $\sim 1.08^{\circ }$, the Moire wave-length is about 15 nm, and
the low energy bandwidth is most significantly suppressed down to about $10$
meV. Meanwhile the local Coulomb interaction was estimated to be of order 10
meV (Ref.\cite{Cao2018Corr}). The experimentalists showed correlated
insulating phases with $\sim 0.31$ meV gap at half-filling of the valence
and conductance bands. Besides, they even observed highly unconventional
superconductivity down to $\sim 1.7$ K near the half-filling\cite{Cao2018SC}.

Nevertheless, twisting is not the only way to Moire superlattice. The
hexagonal boron nitride (hBN) that is commonly used as the substrate for
graphene shares almost the same honeycomb lattice with graphene, but with $%
1.8\%$ larger lattice constant. When the hBN is carefully aligned with a
generic multi-layer graphene, this tiny mismatch of the lattice could induce
a Moire superlattice with Moire wavelength up to $\sim 15$ nm. The band
folding and greatly suppression of kinetic energy would equally give the
possibility to correlated physics. Indeed, not long after the publication by
Pablo's group, the group led by Feng Wang also reports their discovery of
correlated insulating states in the heterostructure of ABC-stacked trilayer
graphene (TLG) over hBN\cite{FengWang}. The choice of multi-layer graphene,
despite its difficulty in fabrication, is on the purpose of suppressing
kinetic energy to the best, as multi-layer results in higher order
energy-momentum dispersion near charge neutral point\cite{FengWang}. It
turns out that the low energy valence bandwidth is also about 10 meV, and
the local Coulomb interaction is also estimated to be of order of magnitude $%
\sim 15$ meV. In their experiment, correlated insulating state is observed
not only at half-filling but also quarter-filling, with $\sim 2.2$ meV gap
at half filling and $\sim 0.5$ meV gap at quarter filling. And the
insulating signal is prominent in valence band but not conductance band.

Now we have two graphene-based Moire superlattice systems in experiments,
both of which have similar Moire wave-length, low energy bandwidth, and
estimated local Coulomb interaction. However, besides the insulating gap
value, the two systems are qualitatively distinct in the relevant band
structure. In TBG, the $C_{2}\mathcal{T}$ protected Dirac cones glue the
valence band and conductance band, and partially filling either bands shows
insulating behavior. In contrast, in TLG/hBN, the Dirac cone is no longer
protected due to the breaking of $C_{2}\mathcal{T}$, therefore separating
the valence band with the conductance band. Moreover, the experiment shows
prominent insulating behavior in partial filling valence band instead of
conductance band, suggesting highly particle-hole asymmetry with respect to
the charge neutral point. While many people argue for a minimal model
capturing the Dirac cones to describe the TBG, it seems that a simpler
single band model for the valence band is sufficient to describe the
correlated physics in TLG-hBN. Simpler band structure sheds light on the
essential correlated physics. Hence we mainly focus on the TLG/hBN system
despite the flooding interests in TBG.

Although the Moire bandstructure of generically twisted multi-layer graphene
over hBN has already been intensively studied in the past ten years\cite%
{Neto,MacDonald2011,Magaud2012,Miller12PRB,WallbankFalko}, the newly
emergent correlated physics such as Mott phases and unconventional
superconductivity is far beyond the description of noninteracting band
theories. To reveal the strongly correlated nature, a simplest minimal model
capturing the only relevant Moire band and the strong interactions is
urgently needed. Indeed, a series of subsequent works concerning this issue
have already been proposed soon after the TBG experiment\cite%
{Senthil,LiangFu,Scalettar,ZhuXiangZhang,Baskaran,Skryabin,Kivelson,FanYang,Ma,PatrickLee,Vafek,Mellado,Cenke,Roy,Das,LiangFuNesting,YiZhuang,ChaoMing,SenthilValleyChern}%
. In a paper by the present authors, the effective bandstructure in TLG/hBN
was calculated by the continuum Dirac model, based on which a weak coupling
minimal model was proposed to describe the low energy valence band\cite%
{ZhuXiangZhang}. Among the variety of recent theoretical works, we were the
first to demonstrate that the half-filled Fermi-surface (FS) of the two
valleys in TLG are exposed to a strong nesting effect and van Hove
singularizes, leading to the IVS order \cite{ZhuXiangZhang}. We also pointed
out that a similar mechanism is likely to happen in TBG system, except that
there would be a triple-Q nesting between valleys instead of single Q
nesting. The triple-Q nesting scenario has been firmly demonstrated by a
series of subsequent works \cite{FanYang,LiangFuNesting,YiZhuang}. From this
weak coupling picture, the Mott features in both TBG and TLG/hBN systems
near half-filling are mainly attributed to the nesting valley FS. However,
the weak coupling nesting scenario is hard to account for both the Mott
insulating states at half-filling and quarter-filling. The fact that a gap
opens whenever one additional electron is added per site points towards a
strong coupling tendency.

In this paper we come to address the issue of quarter-filling in TLG/hBN
from the strong coupling aspects, where the spin-spin repulsion, the
valley-valley repulsion and the Hund's coupling are considered. From the
strong coupling limit we derive the effective spin-valley-exchange
interactions. The interactions contain the most generic terms allowed by U(1)%
$_v\times$SU(2)$_s$ symmetry, including the anti-symmetric
Dzyaloshinsky-Moriya (DM) like interaction. While we derive the effective
model from the TLG-hBN system, this spin-valley model with generic flux
could apply to other graphene-based Moire superlattice. This spin-valley
model is qualitatively distinct from the conventionally studied spin-orbital
model, due to the exceptional appearance of valley-contrasting flux. A very
important consequence is the breaking of sub-valley reflection symmetry that
has severe impact on the Mott state and pairing symmetry. We sketch the
classical phase diagram for varying parameters by minimizing the energy. The
valley-contrasting flux remarkably enlarges the phase space of the
ferro-spin inter-valley 120$^\circ$ order, which shares the same valley
feature as the IVS order from weak coupling theory \cite{ZhuXiangZhang}.
When the filling deviates from exact quarter-filling, the leading pairing
instability is found to be the inter-valley pairing, whose form factor is a
trigonally-warped $(p\pm ip)$-wave. Depending on the sign of the Hund's
coupling, the pairing could favour spin triplet or singlet. The spin singlet
pairing is necessarily chiral and breaks $\mathcal{T}$. Within the spin
triplet channel, the total pairing state could be either chiral
superconductor or helical superconductor protected by time reversal symmetry.

This paper is organized as follows. After brief reviewing the Moire band
structure and minimal model of the TLG/hBN system, we propose the
spin-valley extended Hubbard model in section \uppercase\expandafter{%
\romannumeral2}. In section \uppercase\expandafter{\romannumeral3}, we go to
the strong coupling limit and derive the effective spin-valley-exchange
interactions and discuss its ground state order. In section \uppercase%
\expandafter{\romannumeral4}, we investigate the leading pairing instability
when the Mott insulator is lightly doped. After a brief summary in section %
\uppercase\expandafter{\romannumeral5}, we'll compare our theories with
others and discuss the experimental signals and some further generalization
of our model in the final section \uppercase\expandafter{\romannumeral6}.

\section{Moire band structure and spinful minimal model}

\begin{figure}[t]
\centering
\includegraphics[width=8cm]{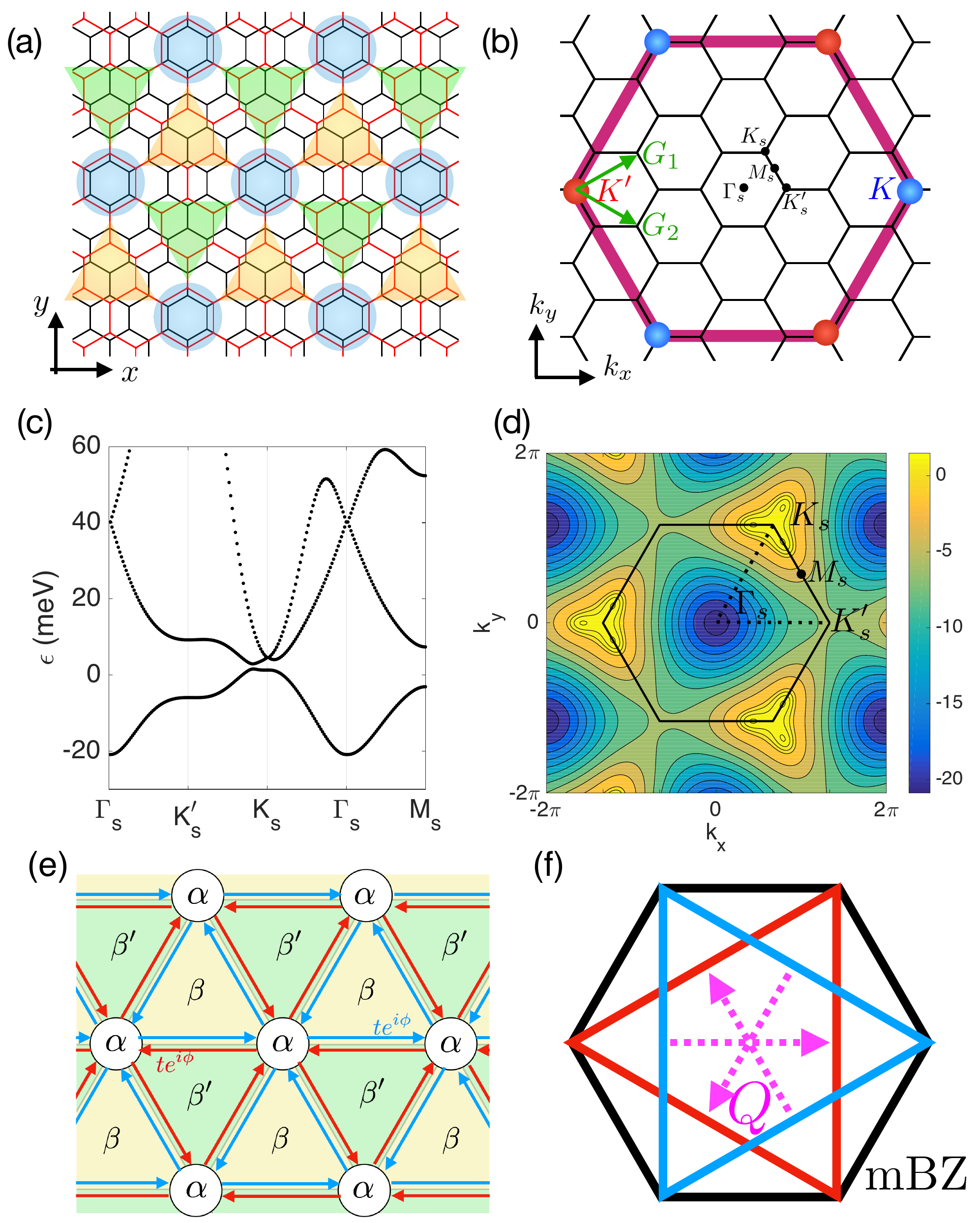}
\caption{(a) Super-lattice formed by TLG (abstracted by a black honeycomb
lattice) and hBN (red honeycomb lattice). For the sake of clearness, we
exaggerate the lattice constant mismatch to $33\%$. The Moire pattern is
composed of three interlaced regions shaded by blue, yellow, and green. The
blue one shows the maximal alignment between the TLG and hBN, while yellow
and green have maximal misalignment and are related by $C_{6}$. (b) The
Brillouin zone of the TLG on the original lattice (marked by the purple
hexagon) is folded into many mini-Brillouin zone by the Moire periodic
potential. (c) Low-energy Moire band structure for the valley $K$ whose
Dirac points are close to $K_{s}$ in the mini-BZ. The Dirac points near $%
K_{s}$ are gapped out by the Moire potential, which is approximated by $%
V_{M}\approx 80$meV. (d) Contour plot of the corresponding valence band near
the CNP in the mBZ (black hexagon). The vicinity of $K_{s}^{\prime }$ hosts
three saddle points where the density of states diverges for this valley
band. Color represents energy in unit of meV. (e) The minimal model features
a valley-contrasting staggered flux. (f) The Fermi surfaces of two valleys
(the red and blue triangles) at half-filling are nested by $Q=(4\protect\pi %
/3,0)$ and its equivalents. }
\label{lattice}
\end{figure}

In this section we briefly review the effective bandstructure and minimal
model before introducing the complete interaction terms. The ABC-stacked TLG
has the same Bravais lattice as in the monolayer graphene. But the electron
and hole touching at zero energy support chiral quasiparticles with $3\pi $
Berry phase, generalizing the low-energy band structure of the monolayer and
bilayer graphene and characterizing a trigonal warped triple Dirac
dispersion in each of the two valleys\cite{KoshinoMcCann}. The hBN also
forms a honeycomb lattice but has a lattice constant about $1.8\%$ larger
than that of the graphene. Thus the heterostructure of TLG and hBN breaks
the original lattice periodicity and there emerges a large scale triangular
Moire super-lattice as shown in Fig.\ref{lattice}a, which contains three
interlaced regions in each Moire unit cell. The TLG/hBN heterostructure
possesses the three-fold rotational symmetry along the $z$-axis $C_{3}$, the
mirror reflection symmetry with respect to the y-z plane $M_{x}$ and the
time reversal symmetry $\mathcal{T}$.

The hBN has a large charge gap $\sim 4.6$ eV and therefore contributes only
a Moire periodic potential to the low energy dynamics in TLG. Stemming from
the difference between boron and nitride atoms, the Moire potential breaks
the protecting symmetry of the Dirac points i.e. sub-lattice symmetry $C_{2}%
\mathcal{T}$ (Ref.\cite{Rappe,Miller12PRB}). Viewed in momentum space, the
Moire periodic potential scatters the low energy valleys repeatedly to form
a Moire reciprocal lattice. Within the Moire mini-Brillouin zone (mBZ), the
dispersion is folded and split into many Moire mini-bands, whose energy
scale decreases from the original bandwidth by orders of magnitude\cite%
{Neto, MacDonald2011, Magaud2012, Miller12PRB, WallbankFalko}. Notice that
as the distance between the two valleys are about $62.5$ times longer than
the Moire wave-vector, the Moire coupling between the two valleys are
exponentially small and can be neglected. As a result, the valleys that were
connected within one band are now well separated and compose two degenerate
Moire mini-bands. In real space, that means the valley degree of freedom
becomes a local orbital in Moire super-lattice, analogous to the physical
spin.

Using the effective two-component Hamiltonian for the TLG \cite%
{KoshinoMcCann}, we have calculated the band structures with the first
harmonic component of Moire scattering potential $V_{M}$ assumed to act only
on the bottom graphene layer\cite{FengWang,ZhuXiangZhang}. The low energy
band dispersion is shown in Fig.~\ref{lattice}c and \ref{lattice}d. Indeed,
the kinetic energy scale is quenched from about 1 eV to around 20 meV. As
the Dirac points are further gapped out by the Moire potential, the valence
band is well isolated. Besides the flatness, there are two most significant
features in this bandstructure. First, the half-filled Fermi-surface (FS) is
subjected to remarkably good nesting instability between the two valleys.
Second, the splitting of Dirac cones leaves triple van Hove points along
zone boundary in the vicinity of $K_{s}^{\prime }$. The stronger Moire
potential, the closer the van Hove points merge towards $K_{s}^{\prime }$.

Upon Fourier transformation, we can parametrize the isolated valence band by
a tight binding model in real space triangular Moire super-lattice. Note
that due to the singleness of the band, we directly Fourier transform the
band dispersion and the hopping integrals obtained in this way is unique,
regardless of the gauge of the Bloch wave-function and the choice of Wannier
orbitals.

\begin{table}[h]
\caption{Parametrize the valence band of valley + with tight binding model
on triangular Moire superlattice. $t_j$ stands for the j-th nearest neighbor
hopping integral. $t_1$ stands for the nearest hopping in x direction while $%
t_2$ stands for the next nearest neighbor hopping in y direction and $t_3$
measures the 3rd nearest neighbor hopping in x direction. The parameters
differ by different $V_M$. All are in unit of meV.}
\label{HopInt}%
\begin{tabular}{ccccccc}
\toprule $V_M$ & 80 & 100 & 200 &  &  &  \\
\colrule $t_1$ & 2.1266 $e^{i 0.1128\pi}$ & 1.7702 $e^{i 0.1209\pi}$ &
0.9013 $e^{i0.1419\pi}$ &  &  &  \\
$t_2$ & 0.1344 $e^{i 0\pi}$ & 0.1160 $e^{i 0\pi}$ & 0.0620 $e^{i 0\pi}$ &  &
&  \\
$t_3$ & 0.0411 $e^{-i0.4560\pi}$ & 0.0129 $e^{-i0.5639\pi}$ & 0.0099 $e^{i
0.2288\pi}$ &  &  &  \\
\botrule &  &  &  &  &  &
\end{tabular}%
\end{table}

As is shown in the Table.~\ref{HopInt}, it turns out that the hopping
integral on the triangular lattice decays very fast: the next-nearest
neighbor hopping is smaller than the nearest neighbor hopping by one order
of magnitude, and the third nearest neighbor hopping is further smaller by
two order of magnitude. Therefore, as a simplest minimal model, we take only
the nearest neighbor hopping, which is estimated to be about $1$ meV in
order to produce a bandwidth of about 10 meV. The most important thing is
the presence of the complex phase of the nearest neighbor hopping, which
breaks the inversion symmetry inside the valley. And since the two valleys
are related by $M_{x}$, the valley rotation symmetry is broken down to the
valley conservation symmetry i.e. SU(2)$_{v}\rightarrow $U(1)$_{v}$. The
complex phase shapes the band structure dramatically and ranges slightly
different depending on the different Moire potential $V_{M}$. However, as
the Moire potential $V_{M}$ grows stronger, the band dispersion acquires an
asymptotic particle-hole symmetry. In fact, when the phase factor becomes $%
\pi /6$, there emerges a particle hole symmetry and the Fermi-surface at
half-filling is exposed to perfect nesting instability.

For simplicity, we'll just consider the ideal phase $\pi /6$ in the
following discussion and argue that it explains the essential physics in the
real materials in the absence of perpendicular electric field. Taking into
account the symmetries $C_{3}$, $M_{x}$, and $\mathcal{T}$ , the hopping
integral distribution in the triangular super-lattice is shown in Fig.~\ref%
{lattice}e. We can see that there exhibits a valley-contrasting staggered
flux in each elementary triangles, and the flux is $\pm 3\phi =\pm \pi /2$
depending on the valley and the triangles. Thus the minimal tight-binding
model for the valence band of the TLG-hBN heterostructure is given by the
Hamiltonian
\begin{equation}
H_{t}=\sum_{r,\nu ,\sigma }\sum_{\mathbf{\delta }}\left( -te^{i\nu \phi
}c_{r+\mathbf{\delta },\nu ,\sigma }^{\dagger }c_{r,\nu ,\sigma
}+h.c.\right) -\mu n_{r,\nu ,\sigma },
\end{equation}%
where $\delta =(1,0)$ and $(-1/2,\pm \sqrt{3}/2)$ are the nearest
neighboring vectors of the primitive unit cell. The band dispersion of
valley $\pm $ is given by $\epsilon _{\pm k}=-2t\sum_{\delta }\cos (k\cdot
\delta +\phi )-\mu $, which varies with the phase $\phi $. In general, the
valley band dispersion is trigonally warped by the nonzero flux that breaks
the sub-valley reflection symmetry $M_{x}\tau _{x}$ or the sub-valley
six-fold rotation $C_{6}\tau _{x}$, along with which shift the van Hove
singularity points.

In the previous paper \cite{ZhuXiangZhang}, we mainly discuss the
instability of FS from the weak coupling scenario, where the perfect nesting
condition leads to a logarithmic divergence of inter-valley susceptibility
and strongly enhance the valley interaction. This justifies why we neglect
the spin interaction but focus on the valley interaction. Now in this paper,
we take a different starting point, namely, the strong coupling limit and
treat the valley and spin interaction on equal footing. This is motivated by
the experimental signature that even the 1/4 filling exhibits insulating
behavior. Given the large Moire unit cell, we assume that only on-site
interactions dominate and neglect the long range interactions. Restricted by
the symmetries, the on-site interactions contain the spin-spin Hubbard
repulsion and valley-valley Hubbard repulsion as well as the Hund's
coupling:
\begin{equation}
\begin{split}
H_{int}& =V\sum_{r,\sigma ,\sigma ^{\prime }}n_{r,+,\sigma }n_{r,-,\sigma
^{\prime }}+U\sum_{r,\nu }n_{r,\nu ,\uparrow }n_{r,\nu ,\downarrow } \\
& -2J_{H}\sum_{r}\left( S_{r,+}\cdot S_{r,-}+\frac{n_{r,+}n_{r,-}}{4}\right)
.
\end{split}%
\end{equation}%
where we define the spin operator on each valley as $S_{r,\nu }$ and
correspondingly we can define the valley-isospin operator on each site in
terms of Pauli matrix as $T_{r,\sigma }$:
\begin{equation}
\begin{split}
& S_{r,\nu }^{a}=\frac{1}{2}\sum_{\alpha ,\beta }c_{r,\nu ,\alpha }^{\dagger
}\sigma _{\alpha \beta }^{a}c_{r,\nu ,\beta }(a=x,y,z), \\
& T_{r,\sigma }^{a}=\frac{1}{2}\sum_{\alpha ,\beta }c_{r,\alpha ,\sigma
}^{\dagger }\tau _{\alpha \beta }^{a}c_{r,\beta ,\sigma }(a=x,y,z).
\end{split}%
\end{equation}%
So the total spin and valley operator on each site is $S_{r}^{a}=\sum_{\nu
}S_{r,\nu }^{a}$ and $T_{r}^{a}=\sum_{\nu }T_{r,\sigma }^{a}$ ($a=x,y,z$).
Note also that the pair hopping term from one valley to another is
neglected. For the usual Hund's coupling, $J_{H}>0$, which guarantees that
the on-site orbital singlet (spin triplet) gains energy from Coulomb
repulsion. However, in general there is no forbidding anti-Hund's coupling $%
J_{H}<0$ that favours the orbital triplet instead.

Let's take a check on the symmetries of this Hamiltonian. The presence of
valley-contrasting flux breaks valley $\text{SU}(2)_{v}$ down to U(1)$_{v}$
and breaks the sub-valley reflection symmetries. The Hund's coupling breaks
the independent spin rotation on each valley leaving a total spin rotation
symmetry $\text{SU}(2)_{+}\times \text{SU}(2)_{-}\rightarrow \text{SU}%
(2)_{s} $. As a result, the strong coupling Hamiltonian should be
generically invariant under U(1)$_{c}\times $ U(1)$_{v}\times $SU(2)$_{s}$.
Moreover, when it comes to the spatial symmetries, the sub-valley reflection
symmetry $M_{x}\tau _{x}$ is broken, as well as the sub-valley six-fold
rotation $C_{6}\tau _{x}$. As we'll show later, the breaking of this
sub-valley reflection symmetry changes the Mott state and pairing symmetry
dramatically.

\section{Strong coupling effective theory at quarter filling}

At quarter filling, there is one electron on each site on average. The limit
$U,V\gg t$ would expel any double occupied states and freeze the charge,
giving rise to a Mott insulator as observed by the experiment. Nevertheless,
the spin and valley degree of freedoms are mobile and can win the energy
from the virtual hopping process. To obtain the leading effective
interaction, let's first consider two neighboring sites, $r$ and $r+\delta $%
. The phase factor of the hopping can then be equivalently treated as
gauging the phase of electron on $r+\delta $: $c_{r,\nu }\rightarrow
c_{r,\nu }$ and $c_{r+\delta ,\nu }\rightarrow c_{r+\delta ,\nu }e^{-i\nu
\phi }$. The 16 possible states of the two electrons can be labeled by their
total spin and valley quantum number$\left. |S,S_{z};T,T_{z}\right\rangle $.
However, the anti-symmetric states have the privilege of virtual hopping and
lowering the energy. The lowered energy can be simply calculated by treating
the hopping terms as perturbation and performing second order perturbation.
The anti-symmetric condition locks the total spin and valley quantum number
for the states with low lying energy. The result is shown in Table~\ref{Eng}%
.
\begin{table}[h]
\caption{Low lying states of the two sites labeled by the spin and valley
quantum number.}
\label{Eng}%
\begin{tabular}{ccc}
\toprule Energy levels & channels & deg \\
\colrule 0 & $\left|S=0;T=0\right>$ \& $\left|S=1;T=1\right>$ & 10 \\
$-\frac{4t^2}{U}$ & $\left.|S=0; T=1, T_z=\pm 1\right\rangle $ & 2 \\
$-\frac{4t^2}{V+J_H}$ & $\left.|S=0; T=1, T_z=0\right\rangle$ & 1 \\
$-\frac{4t^2}{V-J_H}$ & $\left.|S=1; T=0\right\rangle$ & 3 \\
\botrule &  &
\end{tabular}%
\end{table}

By means of projector onto each level labeled by good quantum number we
could immediately write down the effective Hamiltonian as $%
H_{J}=\sum_{r,\delta }h_{r,r+\delta }$, where the local spin-valley-exchange
interactions dressed by flux is given be
\begin{equation}
\begin{split}
& h_{i,j}=\frac{1}{4}\left( J_{1}+2J_{2}-J_{0}\right) S_{i}\cdot S_{j} \\
& +\left[ \left( J_{0}+J_{1}\right) S_{i}\cdot S_{j}+\frac{3J_{0}-J_{1}}{4}%
\right] \left[ T_{i}\cdot \left( e^{iT^{z}2\phi }Te^{-iT^{z}2\phi }\right)
{}_{j}\right] \\
& -2\left( J_{1}-J_{2}\right) \left( S_{i}\cdot S_{j}-\frac{1}{4}\right)
T_{i}^{z}T_{j}^{z},
\end{split}%
\end{equation}%
where we have denoted the energy gain from virtual hopping of intra-valley
spin singlet channel, the inter-valley spin singlet channel and the
inter-valley spin triplet channel respectively as $J_{2}\equiv \frac{4t^{2}}{%
U}$, $J_{1}\equiv \frac{4t^{2}}{V+J_{H}}$ and $J_{0}\equiv \frac{4t^{2}}{%
V-J_{H}}$. The normal valley-exchange interaction is modified by the flux as
\begin{equation}
\begin{split}
& T_{r}\cdot \left( e^{iT^{z}2\phi }Te^{-iT^{z}2\phi }\right) _{r+\delta } \\
& =\left( \text{cos}2\phi \right) T_{r}\cdot T_{r+\delta }+(1-\text{cos}%
2\phi )T_{r}^{z}T_{r+\delta }^{z} \\
& \text{ \ \ }-\left( \text{sin2$\phi $}\right) \left( T_{r}\times
T_{r+\delta }\right) \cdot \hat{z},
\end{split}%
\end{equation}%
which introduces additional easy-plane interaction as well as the
anti-symmetric DM interaction. Therefore, the valley SU(2)$_{v}$ has been
broken, and the DM interaction further breaks the sub-valley reflection
symmetry $M_{x}\tau _{x}$.


Since the exchange interactions are overall repulsive, the system is
expected to exhibit overall anti-ferromagnetism, which can be contributed by
either the spin or the valley. Without the knowledge of the accurate
parameters, we'll try to understand this spin-valley model in a broad range
of parameters. While it is hard to accurately solve the quantum model
exactly, as a first step we treat the model in the classical limit, where
the spin and valley-isospin are treated as vectors and the ground state
energy is minimized by minimizing the energy on each bond.

\begin{figure}[t]
\centering
\includegraphics[width=7.9cm]{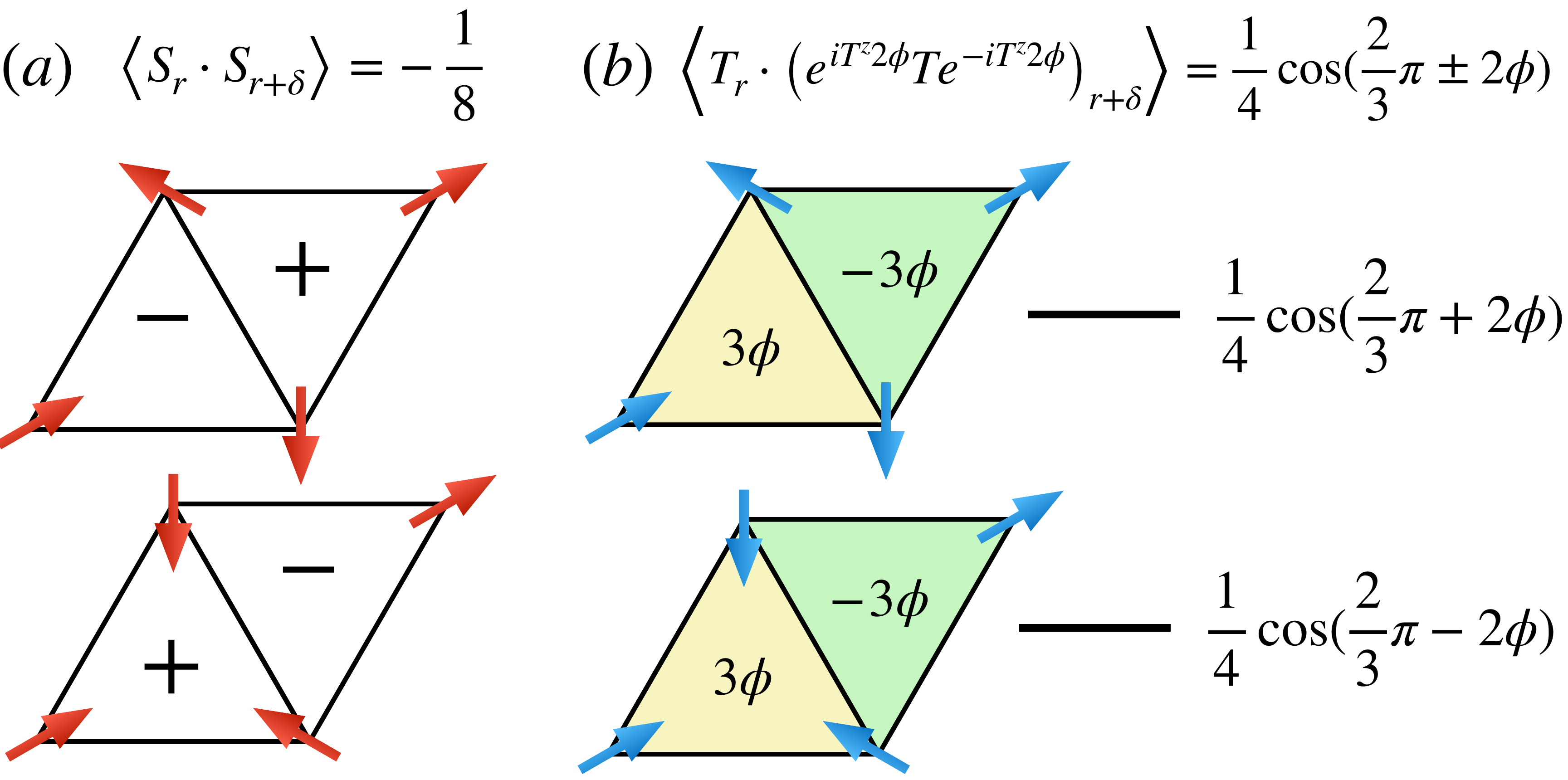}
\caption{(a)The two degenerate spin 120$^{\circ }$ order configurations
related by inversion symmetry. (b)The two valley-isospin 120$^{\circ }$
order configurations related by sub-valley reflection symmetry. As the
sub-valley reflection symmetry is broken by nonzero flux, the degeneracy
between two valley-isospin 120$^{\circ }$ configurations are lifted. }
\label{120order}
\end{figure}

Recall that on a triangular lattice with antiferromagnetic Heisenberg
interaction, the Neel order along $S^{z}$ direction is frustrated by the
lattice, and the dipole spins are compromised to form a classical coplanar
120$^{\circ }$ order \cite{Pierre,Sorella,KunYang}. The spin 120$^{\circ }$
order exhibits alternating spin chirality around the elementary triangle
plaquettes. Therefore this order has two degenerate configurations, that
differ by the spin chirality on a given plaquette, as shown in Fig.~\ref%
{120order}a. For both classical configurations at zero temperature, the
expectation value of exchange term of each bond saturates to the same value:
$\langle S_{r}\cdot S_{r+\delta }\rangle =\frac{-1}{8}$. This is due to the
reflection symmetry $M_{x}$ that relates the two configurations. In
contrast, when it comes to the valley-isospin, the two valley-isospin 120$%
^{\circ }$ configurations are related by the sub-valley reflection symmetry $%
M_{x}\tau _{x}$ instead. Therefore, the breaking of $M_{x}\tau _{x}$ lifts
the degeneracy between the two valley-isospin 120$^{\circ }$ configurations:
\begin{equation}
\left\langle T_{r}\cdot \left( e^{iT^{z}2\phi }\text{$T$}e^{-iT^{z}2\phi
}\right) _{r+\delta }\right\rangle =\frac{1}{4}\cos (\frac{2}{3}\pi \pm
2\phi ),
\end{equation}%
in which the $\pm $ sign depends on the relative sign between the isospin
chirality and flux around a given plaquette, as shown in Fig.~\ref{120order}%
b. In our situation, $\phi =\pi /6$. Hence for the configuration where
positive sign of isospin chirality matches positive sign of flux on a
plaquette, $\langle T_{r}\cdot (e^{iT^{z}2\phi }\text{$T$}e^{-iT^{z}2\phi
})_{r+\delta }\rangle =-1/4$; for the configuration where the sign of the
isospin chirality and flux does not match, $\langle T_{r}\cdot
(e^{iT^{z}2\phi }\text{$T$}e^{-iT^{z}2\phi })_{r+\delta }\rangle =1/8$. In
this sense, the flux stabilizes one of the valley 120$^{\circ }$
configurations but repels the other one.

Meanwhile, the flux also lifts the degeneracy between the planar and Ising
ferromagnetic states of valley-isospin. For convenience, we denote in the
following the energetically favorable valley-isospin 120$^{\circ }$ order as
T-AF, and the spin 120$^{\circ }$ order as S-AF. The Ising ferromagnetism of
valley is abbreviated as T-Fz, which shows expectation
\begin{equation}
\langle T_{r}\cdot (e^{iT^{z}2\phi }\text{$T$}e^{-iT^{z}2\phi })_{r+\delta
}\rangle =\langle T_{r}^{z}T_{r+\delta }^{z}\rangle =1/4.
\end{equation}
On the other hand, the planar ferromagnetic state of valley has
\begin{equation}
\langle T_{r}\cdot (e^{iT^{z}2\phi }\text{$T$}e^{-iT^{z}2\phi })_{r+\delta
}\rangle =1/8, \langle T_{r}^{z}\cdot T_{r+\delta }^{z}\rangle =0
\end{equation}
and this ordered state is abbreviated as T-Fxy. Due to the SU(2)$_{s}$ spin
rotation symmetry, the ferromagnetic state of spin does not discriminate the
Ising and planar ferromagnetism, and can be simply denoted as S-F, which
saturates the expectation value $\langle S_{r}\cdot S_{r+\delta }\rangle =1/4
$. The energy of the several classical order candidates is shown in Table~%
\ref{OrderEng}.
\begin{table}[tbph]
\caption{The energy of the classical orders as combination of the spin and
valley (anti-)ferromagnetism. We draw a comparison between the model with
zero flux and with nonzero flux.}
\label{OrderEng}%
\begin{tabular}{ccc}
\toprule classical orders & energy ($\phi=0$) & energy ($\phi=\pi/6$) \\
\colrule I(S-F, T-AF) & $-\frac{3}{8}J_0$ & $-\frac{1}{2}J_0$ \\
II(S-AF, T-AF) & $-\frac{3}{8}(\frac{J_2}{2}+\frac{J_1}{8}+\frac{5J_0}{8})$
& $-\frac{3}{8}(\frac{J_2}{2}+\frac{5J_0}{6})$ \\
III(S-AF, T-Fz) & $-\frac{3}{8}J_2$ & $-\frac{3}{8}J_2$ \\
IV(S-AF, T-Fxy) & $-\frac{3}{8}(\frac{J_2}{2}+\frac{J_1}{2})$ & $-\frac{3}{8}%
(\frac{J_2}{2}+\frac{3J_1}{8}+\frac{5J_0}{24})$ \\
\botrule &  &
\end{tabular}%
\end{table}

\begin{figure}[t]
\centering
\includegraphics[width=7.9cm]{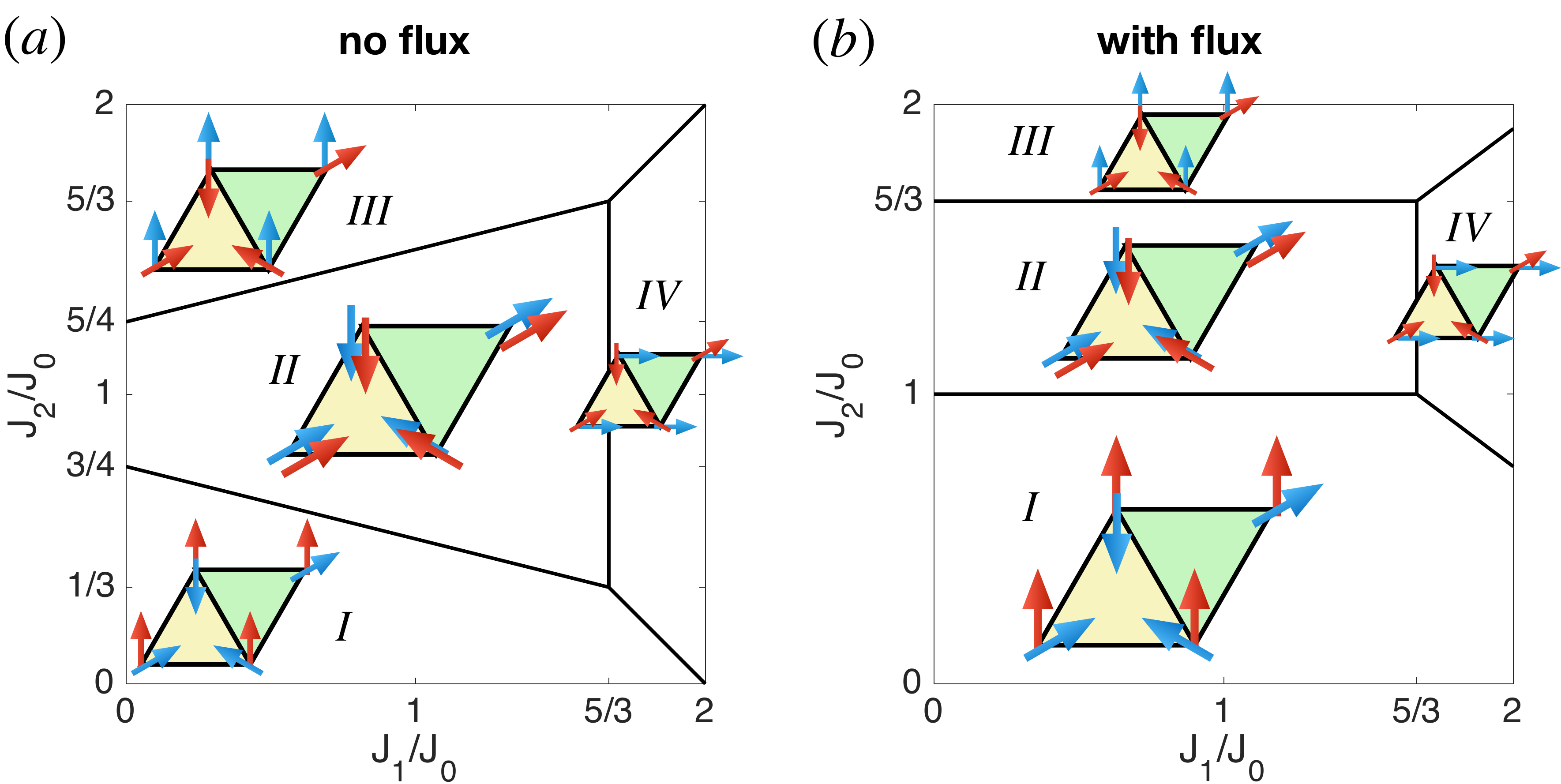}
\caption{ Classical phase diagrams of competing spin-valley orders. (a)
shows the phase diagram in the absence of flux while (b) shows that in the
presence of flux. Both situations exhibit four possible order candidates:
I-the ferro-spin inter-valley 120$^{\circ }$ order (S-F, T-AF), II-the spin
120$^{\circ }$ inter-valley 120$^{\circ }$ order (S-AF, T-AF), III-the spin
120$^{\circ }$ Ising-ferro-valley order (S-AF, T-Fz), the spin 120$^{\circ }$
planar-ferro-valley order (S-AF, T-Fxy). Inside each phase region the inset
shows schematically the order configuration. The blue arrow denotes the
valley isospin while the red arrow denotes the physical spin. Caution that
the inter-valley 120$^{\circ }$ order loses its configuration degeneracy in
the presence of flux and has only one favourable configuration as pinned by
the valley-contrasting staggered flux background. }
\label{Model}
\end{figure}

By minimizing the average bond energy, we sketch the phase diagram shown in
Fig.~\ref{Model}. Before discussing the case with flux, let's first shut
down the flux and look at the more conventional classical phase diagram in
Fig.~\ref{Model}a. The diagonal line ($J_{1}=J_{2}$) in this phase diagram
corresponds to the SU(2)$_{s}\times $SU(2)$_{v}$ symmetric spin-orbital
model \cite{ArovasAuerbach95,Singh98,Affleck2000,Fu-Chun2007,shunqing}.
According to the previous results, the SU(2)$\times $SU(2) symmetric model
under $J_{0}=J_{1}$ exhibits gapless ordered states robust against quantum
fluctuation, partially justifying our assumption of ordering in our
two-dimensional model. The limit of $J_{1}=J_{2}=0$ leaves only $J_{0}$ term
that stabilizes the S-F and T-AF order, and the other limit with $%
J_{1}=J_{2}\gg J_{0}$ stabilizes the S-AF and T-F instead. In between the
more nontrivial gapless phase exists to display the SU(4) symmetry point $%
J_{0}=J_{1}=J_{2}$. The Schwinger boson mean-field approach shows S-AF and
T-AF long range order for this high symmetry model\cite{shunqing}. When $%
J_{1}$ deviates from $J_{2}$, the anisotropy of the orbital space occurs and
leads to different ferromagnetic state of valley-isospin.

Now let's turn on the flux and see what's happening. As shown in Fig.~\ref%
{Model}b, the phase space of the S-F and T-AF is remarkably enlarged. Within
a considerable phase space $J_{1}<5J_{0}/3$, $J_{2}<5J_{0}/3$, the
valley-isospin shows coplanar 120$^{\circ }$ order. This is no wonder when
being reminded that the flux lowers the energy and stabilize the valley
antiferromagnetic state. Near the phase boundaries where more than one
classical orders are degenerate and competing, the quantum fluctuation would
play an important role and possibly lead to nontrivial physics such as spin
liquid. But far away from the phase boundaries, the quantum fluctuation is
supposed to render only tiny correction to the energy but not change the
nature of the ordered state.

Finally, we give a bit comment on the finite temperature behavior. As the
system is a clean 2-dimensional lattice, the Mermin-Wagner theorem defies
continuous symmetry breaking under any finite temperature. Therefore, the
thermal fluctuation would resist the spontaneous breaking of spin SU(2)$_{s}$
symmetry and the formation of spin long range order. Instead, the spin order
parameter fluctuates in real space and can be stabilized by small magnetic
field that explicitly breaks the symmetry\cite{ChaoMing}. Nevertheless, the
breaking of valley U(1)$_{v}$ symmetry can occur through the
Kosterlitz-Thouless transition under finite temperature, establishing the
quasi-long range valley order.

\section{Inter-valley pairing instability}

When the filling deviates slightly from quarter filling i.e. doping the Mott
insulator, the effective Hamiltonian would additionally involves a hopping
term of the charge carriers under occupancy constraint:
\begin{equation}
H_{\text{eff}}=\mathcal{P}\left( H_{t}+H_{J}\right) \mathcal{P}
\end{equation}%
where $\mathcal{P}=\prod_{r}\frac{1}{6}n_{r}(2-n_{r})(3-n_{r})(4-n_{r})$ is
the projector that projects onto the single or null occupancy on each site.
The spin-valley-exchange interaction is under the projection of single
occupation.

To reduce redundancy we can decompose the fermion degree of freedom into the
holon and spin-valleyon $c_{r,\nu ,\sigma }=h_{r}^{\dagger }f_{r,\nu ,\sigma
}$, under the constraint $h_{r}^{\dagger }h_{r}+\sum_{\nu ,\sigma }f_{r,\nu
,\sigma }^{\dagger }f_{r,\nu ,\sigma }=1$. In this way, after some simple
derivation the spin-valley-exchange interaction under constraint $\mathcal{P}%
H_{J}\mathcal{P}$ can be exactly expressed in terms of the spin-valleyon
without redundancy. We can decompose the exchange interaction into 6
anti-symmetric pairing channels which are energetically favourable than the
other 10 symmetric pairing channels:
\begin{equation}
\begin{split}
\mathcal{P}H_{J}\mathcal{P}=& -\frac{1}{4}\sum_{r,\delta }\left\{ J_{2}\vec{%
\Delta}_{2}^{\dagger }(r,r+\delta )\cdot \vec{\Delta}_{2}(r,r+\delta )\right.
\\
& \left. +J_{1}\Delta _{1}^{\dagger }(r,r+\delta )\Delta _{1}(r,r+\delta
)\right. \\
& \left. +J_{0}\vec{\Delta}_{0}^{\dagger }(r,r+\delta )\cdot \vec{\Delta}%
_{0}(r,r+\delta )\right\} ,
\end{split}%
\end{equation}%
where the spin singlet intra-valley channel, spin singlet inter-valley
channel and spin triplet inter-valley channel are denoted respectively as:
\begin{equation}
\begin{split}
& \vec{\Delta}_{2}(i,j)\equiv \psi _{j}e^{-i\tau _{z}\phi }\sigma _{y}\tau
_{y}\left( \tau _{x},\tau _{y}\right) \psi _{i}, \\
& \Delta _{1}(i,j)\equiv \psi _{j}e^{-i\tau _{z}\phi }\sigma _{y}\tau
_{x}\psi _{i}, \\
& \vec{\Delta}_{0}(i,j)\equiv \psi _{j}e^{-i\tau _{z}\phi }\vec{\sigma}%
\sigma _{y}\tau _{y}\psi _{i}.
\end{split}%
\end{equation}%
The spin-valleyon basis is compactly expressed as $\psi _{r}\equiv \left(
\begin{array}{cccc}
f_{r,+,\uparrow } & f_{r,+,\downarrow } & f_{r,-,\uparrow } &
f_{r,-,\downarrow }%
\end{array}%
\right) ^{T}$. The above expression is exact without approximation, as long
as the particle constraint is rigorously kept.

\begin{figure}[t]
\centering
\includegraphics[width=7.9cm]{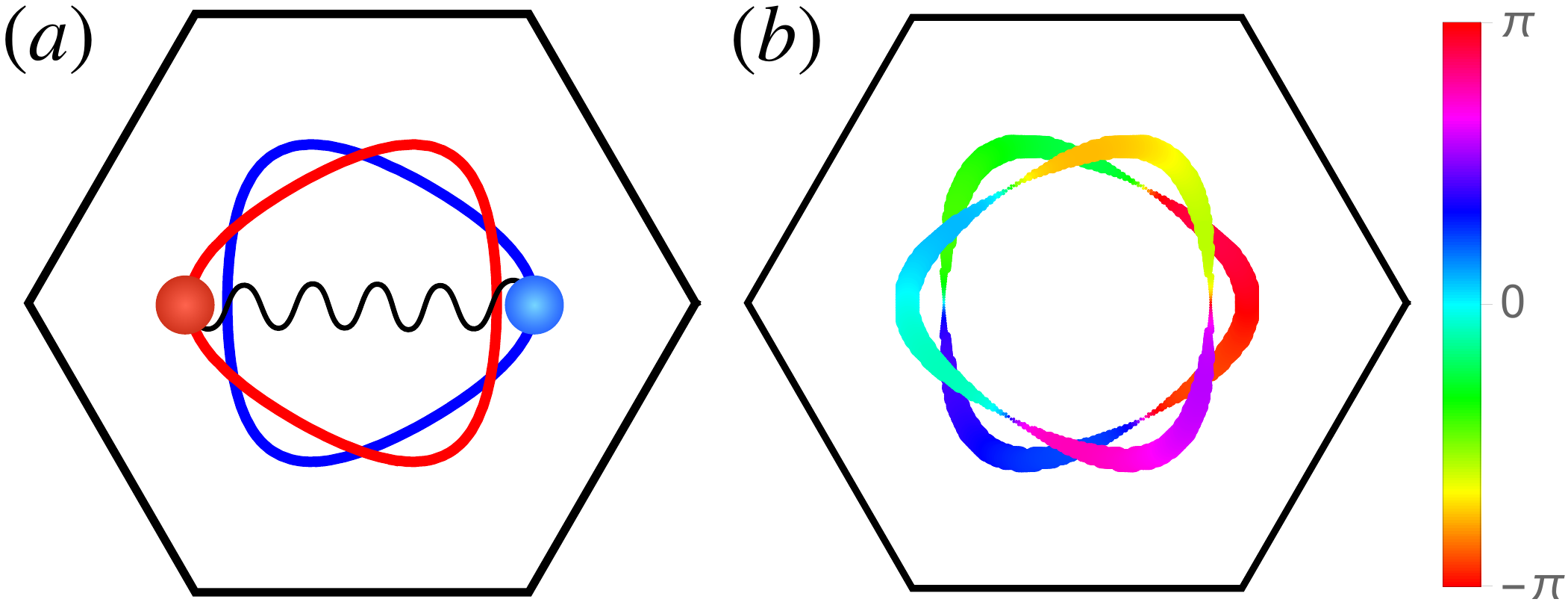}
\caption{(a) The blue and red lines show the FS of valley + and valley -
respectively, near 1/4 filling ($\protect\mu \approx -1.58\tilde{t}$).
Effective valley exchange interaction within spin triplet space induces
inter-valley pairing between opposite valleys and momenta, indicated by the
wavy line. (b)The pairing gap functions of $D_{k}$ are shown on the FS in
(a). The line width is proportional to the gap magnitude, while the color
stands for the pairing phase. It is visualized that the gap minima are
located close to $\mp Q/2$ for valley $\pm $, and pairing phase on both FS
winds +2$\protect\pi $ counter-clockwise. Note that the gap minima are
finite.}
\label{FS&Gap}
\end{figure}

For further discussion we're going to do the approximation and treat the
holon in mean-field level $\langle h\rangle =\langle h^{\dagger }\rangle =%
\sqrt{x}$, with $x$ being the charge carrier density away from one quarter.
Thereby we obtain a renormalized kinetic hopping term for the spin-valleyon:
\begin{equation}
H_{t}\rightarrow \int_{k}\sum_{\nu ,\sigma }\left( \tilde{\epsilon}_{\nu
k}-\mu \right) f_{k,\nu ,\sigma }^{\dagger }f_{k,\nu ,\sigma },
\end{equation}%
in which the hopping amplitude is renormalized by the charge carrier
density: $t\rightarrow \tilde{t}=tx$. The kinetic term determines the
spin-valleyon FSs, which have spin degeneracy but differ by the valleys. The
FS of two valleys are inversion-related triangular warped pocket filling
about 1/4 of the BZ (Fig.~\ref{FS&Gap}a). Due to the sub-valley reflection
symmetry breaking, the single valley FS lacks inversion symmetry and
therefore it is frustrating for two spin-valleyons on the same valley FS to
form a Cooper pair with constant center-of-mass momentum. Therefore we argue
that it is difficult for the intra-valley pair to condense. In the following
we mainly compare the spin singlet and spin triplet inter-valley pairing
channels.

Before introducing the pairing order parameter, we can rewrite the
interaction in the pairing hopping form in momentum space:
\begin{equation}
\begin{split}
& \mathcal{P}H_{J}\mathcal{P}\rightarrow J_{1}\eta _{k,k^{\prime }}\left(
\psi _{k,+}^{\dagger }\sigma _{y}\psi _{-k,-}^{\dagger }\right) \left( \psi
_{-k^{\prime },-}\sigma _{y}\psi _{k^{\prime },+}\right) \\
& +\int_{k,k^{\prime }}J_{0}\eta _{k,k^{\prime }}\left( \psi _{k,+}^{\dagger
}\sigma _{y}\vec{\sigma}\psi _{-k,-}^{\dagger }\right) \cdot \left( \psi
_{-k^{\prime },-}\vec{\sigma}\sigma _{y}\psi _{k^{\prime },+}\right) ,
\end{split}%
\end{equation}%
where the hopping form factor $\eta _{k,k^{\prime }}\equiv -\sum_{\delta
}\cos (k\cdot \delta +\phi )\cos (k^{\prime }\cdot \delta +\phi )$. From
this we can see that the interactions fall in two competing pairing
instability, the spin singlet pairing driven by $J_{1}$ and the spin triplet
channel driven by $J_{0}$ respectively. They share the same form factor and
therefore the outcome only depends on $J_{1}/J_{0}$. If $J_{1}>J_{0}$, the
dominant instability is the spin singlet pairing, while if $J_{1}<J_{0}$,
the dominant instability is the spin triplet pairing. When $J_{1}=J_{0}$,
there emerges SO(4) symmetry and the spin singlet pairing is degenerate with
the spin triplet pairing. In the following we'll be limited inside either
one of the spin channel and discuss the spatial part of the pairing symmetry
determined by the orbital form factor.

The pair hopping interaction is attractive in long range but repulsive in
short range: $\eta _{k,k^{\prime }}<0$ $(k\simeq k^{\prime })$ and $\eta
_{k,k^{\prime }}>0$ $(k\simeq -Q-k^{\prime })$. Therefore the Cooper pair
condensate must change sign across the FS when hopping from $k$ to $-Q-k$ so
as to circumvent the repulsion and gain energy. In this sense, $s$-wave is
unflavored. Besides trivial $s$-wave, the $C_{3}$ symmetry allows another
two chiral representations with angular momentum $\pm 1$. The hopping
interaction can be decomposed into three orthogonal channels according to
the $C_{3}$ representations:
\begin{eqnarray}
\eta _{k,k^{\prime }} &\equiv &-\frac{1}{3}\left( D_{k}D_{k^{\prime }}^{\ast
}+D_{k}^{\ast }D_{k^{\prime }}+S_{k}S_{k^{\prime }}\right) , \\
S_{k} &\equiv &\sum_{\delta }\cos (k\cdot \delta +\phi ),  \notag \\
D_{k} &\equiv &\cos \left( k_{x}+\phi \right) +e^{i2\pi /3}\cos \left(
k_{x}/2-\sqrt{3}k_{y}/2-\phi \right)   \notag \\
&&+e^{-i2\pi /3}\cos \left( k_{x}/2+\sqrt{3}k_{y}/2-\phi \right) .
\end{eqnarray}%
The first channel is $s$-wave that can be excluded, and the latter two are $%
\mathcal{T}$ related chiral representations. The chiral representation
indeed satisfies the sign changing condition across the FS and are
energetically favorable. Under $T_{c}$, the Cooper pair would spontaneously
pick up one of the chiral form factor.

\begin{figure}[t]
\centering
\includegraphics[width=7.9cm]{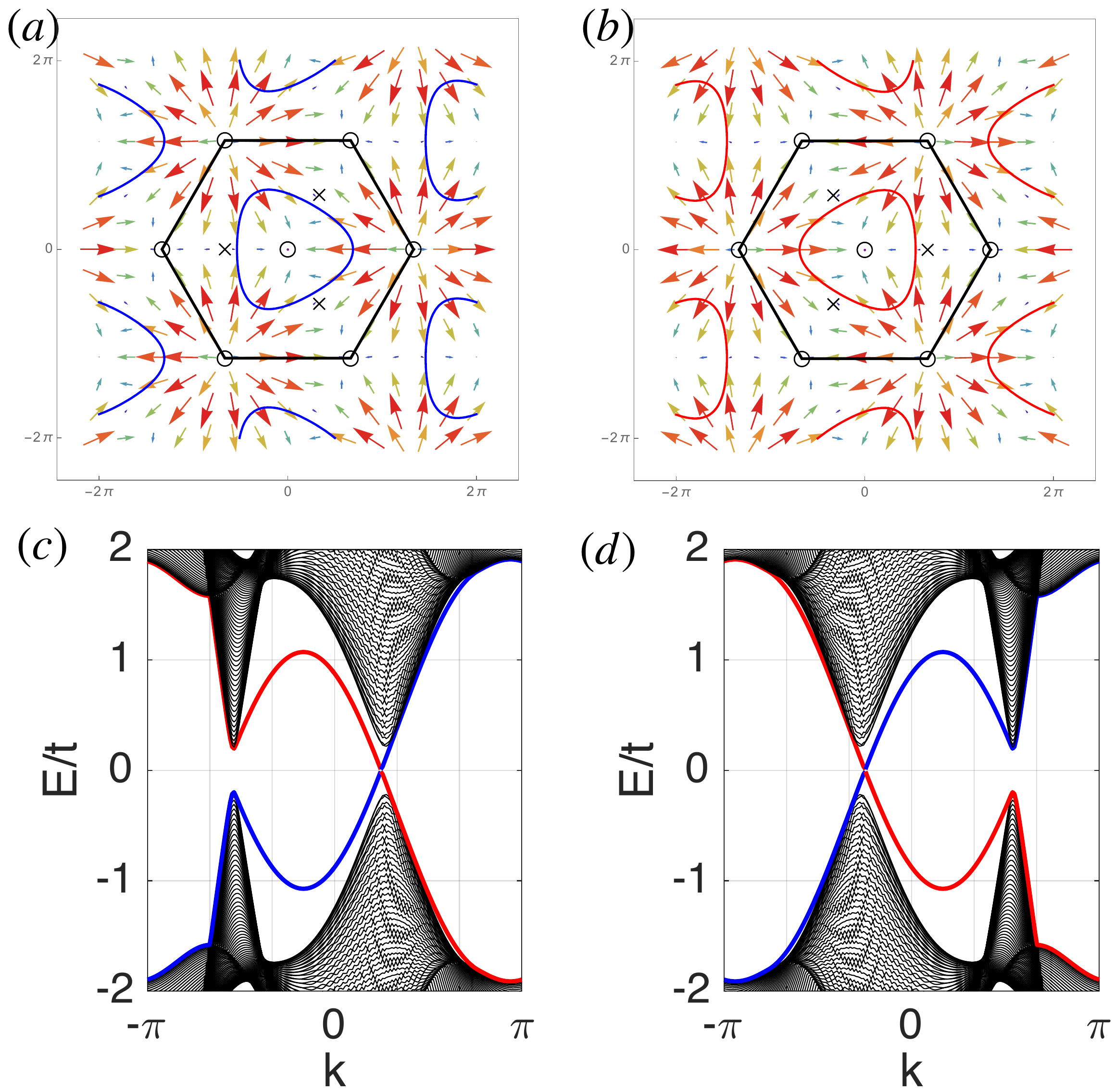}
\caption{(a) and (b) show the distribution of $\vec{d}_{k}$ vector defined
by form factor $D_{k}$ in mBZ, for valley + (left figure) and valley -
(right figure) respectively. The trigonally-warped FS is highlighted by
blue/red lines for valley $\pm $ respectively. The black hexagon marks the
first mBZ. The small black circles inside the figure denote vortex cores
while the black crosses denote anti-vortex cores, both of which are zeros of
the pairing condensate. The vortex cores are located on $\Gamma _{s}$, $%
K_{s} $ and $K_{s}^{\prime }$. The anti-vortex cores reside on $-Q/2$ i.e. $%
(-2\protect\pi /3,0)$ and its $C_{3}$ counterparts. The FSs are fully gapped
and the pairing condensate winds $2\protect\pi $ around them. (c) and (d)
show the Bogoliubov spectra for one of the spin triplet channel being placed
on an open cylinder with smooth edges, for valley + and valley -
respectively. The momentum $k$ follows the edges along $x$ direction. The
blue and red line denote the chiral edge modes on the two edges
respectively, indicating a chiral Bogoliubov edge mode. The chemical
potential $\protect\mu \approx -1.58\tilde{t}$ to ensure filling near 1/4
filling, and the pairing order parameter is chosen as $\Delta =0.5\tilde{t}$
for a clearer demonstration, whose value does not affect the topology.}
\label{PairForm}
\end{figure}

We can examine the pairing symmetry associated with $D_k$ more carefully.
Near the mBZ center $\Gamma _{s}$, the pairing form factor can be expanded
as $D_{k}=-\frac{3}{4}\left( k_{x}+ik_{y}\right) +O\left( k^{2}\right) $,
which contains the $p+ip$ pairing form to the leading order. Indeed, we
investigate the pairing gap on FSs close to quarter filling (Fig.\ref{FS&Gap}%
b), which exhibits amplitude anisotropy and winding of phase by $2\pi $. The
gap maxima are located near the FS corners while the gap minima are at the
middle point of each arc of the near-triangle-shaped FS. To gain a better
insight into this, we explicitly map out the vector field of the complex
pairing form factor $D_{k}$ in the mBZ: $\vec{d}_{k}=\Delta (\text{Re}D_{k},%
\text{Im}D_{k})$. As shown in Fig.\ref{PairForm}, in each mBZ there are
three vortex cores residing on the $\Gamma _{s}=(0,0)$, $K_{s}=(-4\pi /3,0)$%
, and $K_{s}^{\prime }=(4\pi /3,0)$ respectively. Besides, three anti-vortex
cores are located at $-Q/2$, i.e. $(-2\pi /3,0)$ and its $C_{3}$
counterparts. These anti-vortices results in sign changing from $k$ to $-k-Q$%
. The hole doped valley FSs avoid all the zeros and hence are fully gapped,
but each FS encloses single vortex core residing at $\Gamma _{s}$, which
explains the phase winding by $2\pi $. This indicates that it is
adiabatically equivalent to the chiral $p+ip$ pairing condensate, except
that the form factor is trigonally warped by the flux. The mean-field
superconducting Hamiltonian can be easily proved to yield a Bogoliubov
de-Gennes topological Chern number for each valley:
\begin{equation}
C_\nu=\frac{1}{4\pi }\int_{k}\hat{h}_{k}\cdot (\partial _{x}\hat{h}%
_{k}\times \partial _{y}\hat{h}_{k})=1,
\end{equation}
where $\hat{h}_{k}$ is the unit vector of $\vec{h}_{k}\equiv
(d_{k}^{x},d_{k}^{y},\tilde{\epsilon}_{k})$, and $\tilde{\epsilon}_{k}$ is
the renormalized effective kinetic term.

To see that the topological state does support gapless edge modes, we
perform exact diagonalization for the pairing state placed on a cylinder
with smooth edges. As is shown in Fig.~\ref{PairForm}c and \ref{PairForm}d,
the Bogoliubov spectra of Bogoliubov quasi-particles of both valleys support
chiral gapless Bogoliubov edge modes.

The above discussion is limited in a selected channel in the spin space. Now
we come back to address the spin space part of the pairing symmetry. There
are three possible scenarios depending on the parameters.

1. When $J_{1}>J_{0}$, the spin singlet pairing is more energetically
favourable. Since there is only one component in the spin singlet channel,
it is necessarily a chiral state. We could introduce the chiral spin singlet
pairing order
\begin{equation}
\Delta _{chiral}^{s}\equiv \frac{-J_{0}}{3}\int_{k}\left\langle \psi
_{-k,-}\sigma _{y}\psi _{k,+}\right\rangle D_{k}^{\ast },
\end{equation}%
to decouple the interaction to a mean-field pairing term
\begin{equation}
H_{J}^{\text{MF}}=\Delta _{chiral}^{s}\int_{k}D_{k}\left( f_{k,+,\uparrow
}^{\dagger }f_{-k,-,\downarrow }^{\dagger }-f_{k,+,\downarrow }^{\dagger
}f_{-k,-,\uparrow }^{\dagger }\right) +h.c.
\end{equation}%
The edge supports a chiral spinful complex fermion mode associated with
total BdG Chern number $C=4$. The spin SU(2)$_{s}$ symmetry is respected in
this case.

2. When $J_{1}<J_{0}$, the spin triplet pairing is more energetically
favourable. Different from the spin singlet case, the spin triplet pairing
has multi-component and allows room for two $\mathcal{T}$-related pairing
condensates. Therefore it falls into two degenerate situations.

First, the chiral spin triplet pairing state. We could introduce the vector
order parameter in the SU(2) spin triplet space for the chiral spin triplet
pairing:
\begin{equation}
\vec{\Delta}_{chiral}^{t}\equiv \frac{-J_{1}}{3}\int_{k}\left\langle \psi
_{-k,-}\vec{\sigma}\sigma _{y}\psi _{k,+}\right\rangle D_{k}^{\ast },
\end{equation}%
which corresponds to the chiral pair condensate in real space
\begin{equation}
-\frac{J_{\text{S1T0}}}{2}\left\langle \psi _{r+\delta }e^{-i\tau _{z}\phi }%
\vec{\sigma}\sigma _{y}\tau _{y}\psi _{r}\right\rangle =\vec{\Delta}%
_{chiral}^{t}e^{i\frac{2}{3}\pi \delta }.
\end{equation}
Note that $\vec{\Delta}_{chiral}$ is assumed to be real vector so that all
directions can be related by SU(2)$_{s}$ rotation. In contrast, when beyond
real vector, an example like $\vec{\Delta}_{chiral}^{t}=\Delta
_{chiral}^{t}\left( \frac{1}{2},\frac{i}{2},0\right) $ fails to gap out the
spin up FS and is not the most energetically favourable choice. Below $T_{c}$
the vector order parameter spontaneously picks one direction and breaks SU(2)%
$_{s}\rightarrow $U(1)$_{s}$, meanwhile the chiral pair condensate breaks
time reversal symmetry $\mathcal{T}$. To illustrate the mean-field
Hamiltonian after superconductivity sets in, we might as well pick up a
direction like $\vec{\Delta}_{chiral}^{t}=\Delta _{chiral}^{t}(0,1,0)$ and
picks up the $D_{k}$ without loss of generality:
\begin{equation}
H_{J}^{\text{MF}}=\Delta _{chiral}^{t}\int_{k}D_{k}\left( f_{k,+,\uparrow
}^{\dagger }f_{-k,-,\uparrow }^{\dagger }+f_{k,+,\downarrow }^{\dagger
}f_{-k,-,\downarrow }^{\dagger }\right) +h.c.
\end{equation}%
This mean-field pairing term describes a chiral pair condensate for both
spin up and spin down pairing. The edge supports a spin degenerate complex
fermion edge mode, corresponding to the total BdG Chern number $C=4$.

Second, there is the spin triplet helical pairing state. When the orbital
form factors of the Cooper pairs with opposite spin are counter-chiral and
related by $\mathcal{T}$, the $\mathcal{T}$ is preserved. Without loss of
generality, we could write down the helical order parameter up to a choice
of spin basis:
\begin{equation}
\Delta _{helical}=\frac{-2J_{1}}{3}\int_{k}\left\langle f_{-k,-,\uparrow
}f_{k,+,\uparrow }\right\rangle D_{k}^{\ast },
\end{equation}%
and the interaction term is decoupled after superconductivity sets in:
\begin{equation}
H_{J}^{\text{MF}}=\Delta _{helical}\int_{k}D_{k}f_{k,+,\uparrow }^{\dagger
}f_{-k,-,\uparrow }^{\dagger }+D_{k}^{\ast }f_{k,+,\downarrow }^{\dagger
}f_{-k,-,\downarrow }^{\dagger }+h.c
\end{equation}%
which is degenerate with the chiral candidate due to a fictitious symmetry
that mirror reflects the momentum of only the spin down electrons \cite%
{Cenke}. In this case, the BdG Chern numbers of different spin components
are opposite so that the total BdG Chern number is zero but there exists a
spin BdG Chern number $C_{s}=4$ in the presence of $\mathcal{T}$.
Correspondingly, the edge supports a helical complex fermion mode: the spin
up component propagates counter-clockwise with the spin down component.

All in all, the relative strength between $J_0=4t^2/(V-J_H)$ and $%
J_1=4t^2/(V+J_H)$ depends on the sign of Hund's coupling. To conclude,
depending on the sign of the Hund's coupling, the favourable pairing state
can be spin singlet or triplet, as stabilized by $J_1$ or $J_0$. Within the
spin triplet channel the pairing could fall into chiral or helical
degenerate candidates, as shown in Table.~\ref{PairingSym}.
\begin{table}[h]
\caption{Representative candidates of the inter-valley pairing symmetry. }
\label{PairingSym}%
\begin{tabular}{cccc}
\toprule condition & type & pairing form & symmetries \\
\colrule \multirow{2}{*}{ $J_H>0$} & chiral & $(p\pm
ip)_{\uparrow\uparrow}+(p\pm ip)_{\downarrow\downarrow}$ & U(1)$_v\times$
U(1)$_s$ \\
& helical & $(p\pm ip)_{ \uparrow\uparrow}+(p\mp ip)_{\downarrow\downarrow}$
& U(1)$_v\times \mathcal{T}$ \\
$J_H<0$ & chiral & $(p\pm ip)_{\uparrow\downarrow}-(p\pm
ip)_{\downarrow\uparrow}$ & U(1)$_v\times$ SU(2)$_s$ \\
\botrule &  &  &
\end{tabular}%
\end{table}

\section{Discussion and Summary}

\begin{figure}[h]
\centering
\includegraphics[width=7.9cm]{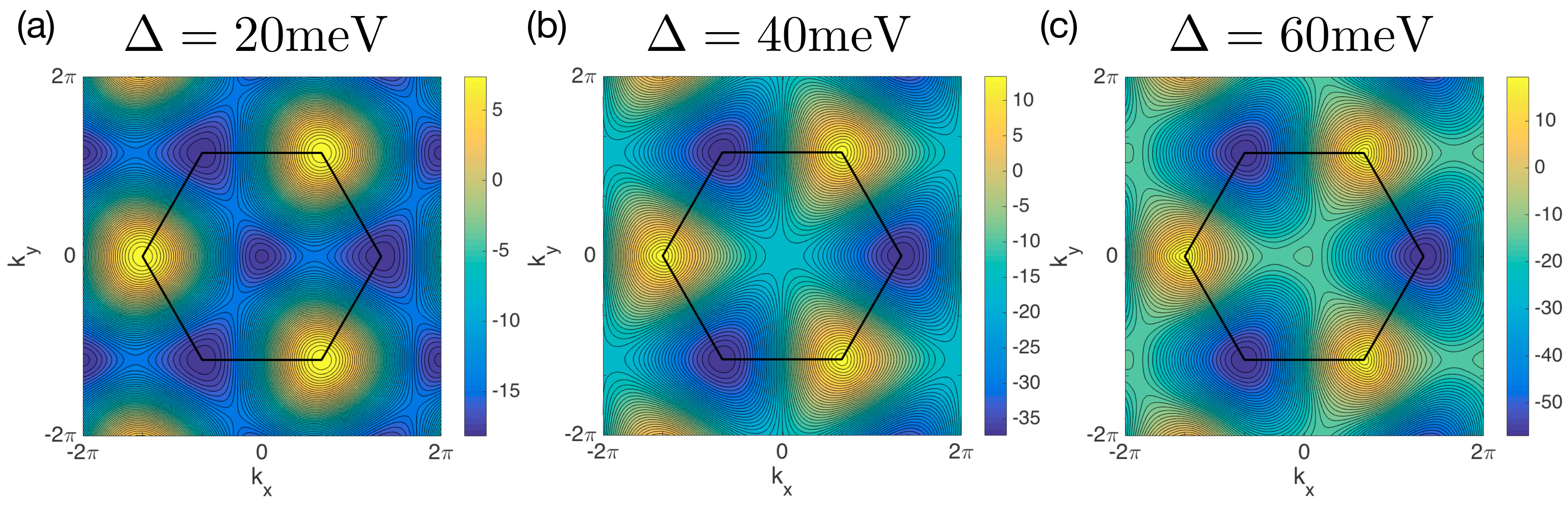}
\caption{Valence band structure in the presence of nonzero perpendicular
electric field $\Delta$ and $V_M$=80 meV. (a) $\Delta$=20 meV, (b) $\Delta$%
=40 meV, (c) $\Delta$=60 meV. The most important feature is the moving of
the van Hove points with increasing $\Delta$.}
\label{GateTuning}
\end{figure}

The perpendicular electric field is not explicitly considered in this paper
by far. However, in the graphene-based multi-layer-hBN system, it is very
easy and natural in experiment to tune this electric field and change the
band structure completely. In fact, as are shown in Fig.~\ref{GateTuning},
the valence band structure in the presence of generically nonzero
perpendicular electric field essentially manifest in the moving of van Hove
points, which can be captured by the flux changing as shown in Table.~\ref%
{VaryFlux}.

\begin{table}[h]
\caption{Parametrize the valence band of valley + with tight binding model
on triangular Moire superlattice. The parameters differ by different
perpendicular field $\Delta$. All are in unit of meV. $V_M$=80meV.}
\label{VaryFlux}%
\begin{tabular}{cccc}
\toprule $\Delta$ & 20 & 40 & 60 \\
\colrule $t_1$ & 2.8648 $e^{i 0.3588\pi}$ & 4.9168 $e^{i 0.4887\pi}$ & 7.0883%
$e^{i 0.5301\pi}$ \\
$t_2$ & 0.1385 $e^{i 0\pi}$ & -0.1053 $e^{i 0\pi}$ & -0.4166 $e^{i 0\pi}$ \\
$t_3$ & 0.1526 $e^{i 0.1561\pi}$ & 0.1719 $e^{i 0.3067 \pi}$ & 0.1660$e^{-i
0.9297\pi}$ \\
\botrule &  &  &
\end{tabular}%
\end{table}

As we've elaborated in the previous and this paper, the valley-contrasting
flux is crucial in determining the correlated physics. Actually, the flux
changing would modify the spin-valley exchange interaction and therefore
could result in different spin-valley orders for the Mott insulators in a
rich phase diagram. Moreover, the flux changing could also change the
leading pairing symmetry dramatically. For example, when the flux is
gradually turned off, the anti-vortices of the pair condensate would
approach towards the $\Gamma_s$ across the FS, which drives a topological
phase transition from $p\pm ip$ to $d\mp id$ pairing symmetry with $C=\mp2$
instead. Actually, the $d\mp id$ pairing symmetry is much more commonly
studied in the triangular lattice driven by repulsive interaction\cite%
{QiangHua, Ziqiang, ZiYang}. We note that the exceptional possibility of $%
p\pm ip$ pairing symmetry in this system is due to the nonzero
valley-contrasting flux. The flux explicitly breaks the sub-valley
reflection symmetry and mixes the Cooper pairs with different parity.

There are some possible experimental consequences of our Mott insulating
orders and the topological superconductivity. First, the Mott states at 1/4
filling is likely to exhibit the inter-valley 120$^{\circ }$ order. This
order entails valley coherence, which may be detected by an optical
experiment as the two valleys are in correspondence to the left and right
circular-polarized light. Besides, concerning the orbital character of the
valley, a spatial charge modulation would occur in the microscopic graphene
lattice. As the two valleys in the original BZ are located on the opposite
corners, the momentum interference induces $\sqrt{3}a\times \sqrt{3}a$
charge pattern in the microscopic graphene lattice. Second, since the $p+ip$
pairing superconductivity is degenerate with that of $p-ip$, in realistic
material there are likely to form domains between these two superconducting
phases below the critical temperature. While the domain is supposed to show
fully superconducting gap, on the domain walls there are expected to be
topologically protected gapless fermion modes. These signatures are amenable
to STM probe. On the other hand, our theory exhibits clear pairing gap
anisotropy, which may possibly be confirmed by laser ARPES measurement with
high resolution.

Compared with some recent works related with the Mott states or
superconductivity of TLG/hBN Moire superlattice \cite%
{Cenke,Senthil,SenthilValleyChern,SenthilTLG}, our theory is different from
theirs in the following aspects. While Xu and Balents study the pairing
symmetry driven by SU(4)-symmetric spin-valley-exchange interaction in the
presence of Hund's coupling, our strong coupling model is shown to feature
valley-contrasting flux and SU(2)$_{s}\times $U(1)$_{v}$ exchange
interaction instead. The breaking of sub-valley reflection symmetry is the
crucial ingredient of our theories, which leads to totally different pairing
symmetries. Senthil et al \cite{Senthil} also presented a valley-contrasting
flux model and discussed its strong coupling effective model, but they did
not specify and discuss the flux effects. Unlike ours, their leading
spin-valley exchange interaction preserves SU(4) symmetry and therefore
would not yield our result. In Senthil's later paper \cite%
{SenthilValleyChern}, he and his coworkers show that the Moire valence band
could have nonzero valley Chern number. They argue SU(4) ferromagnetism and
each Hubbard band is spin-polarized and valley-polarized, which leads to
interaction-driven valley Chern insulators at quarter and half filling.
However, as we've shown in this paper, the energetics favours the
antiferro-valley order instead of the ferro-valley order at 1/4 filling for
quite a large phase region. In this sense, at 1/4 filling the
valley-filtered band might not be fully filled to yield quantized Berry
phase. Our predicted topological superconductivity upon doping is not
related with the valley Chern number. In Senthil's most recent paper \cite%
{SenthilTLG}, by the procedure of Wannier orbital optimization, they convert
the screened Coulomb interaction into the local interactions in tight
binding model, which feature a relatively strong nearest neighbor Hund's
coupling. Hence, they give a more concrete demonstration of how the system
exhibits ferro-magnetism. But this mechanism of ferro-magnetism seems
sensitive with the parameters. In fact, their effective band calculation
neglects the $\gamma _{2}$ and $\gamma _{4}$ terms, which are small but
yield qualitatively different band-structure according to our calculation.
So the accurate parameters and the phase regime still remain an open
question that need experimentalists to resolve.

To summarize, we start from a strong coupling theory which features
valley-contrasting staggered flux and spin-valley interactions. The model
respects a minimal U(1)$_c\times$U(1)$_v\times$SU(2)$_s$ internal symmetry.
Under large U, V limit, we derive the effective model for the system near
quarter filling. The effective spin-valley interactions include the DM
interaction that breaks sub-valley reflection symmetry. We sketch a
classical phase diagram by minimizing the energy per bond. The flux plays
the role of enhancing the ferro-spin inter-valley $120^\circ$ order. While
the spin could exhibits anti-ferromagnetism or ferro-magnetism depending on
the parameters, in a remarkable realistic phase space, the valley shows 120$%
^\circ$ order. This order is stabilized by the flux modified valley exchange
interaction, and shares the same feature as the inter-valley spiral order
arising from the valley FS nesting.

Upon doping, the dominant pairing instability is shown to be the topological
pairing state in inter-valley channel. Depending on the sign of the Hund's
coupling, the pairing state could favour spin triplet or spin singlet,
despite the same pairing form factor. The form factor is a trigonally warped
$p\pm ip$-wave, distinct from the $d\mp id$ pairing state commonly studied
in triangular lattice with antiferromagnetic interaction. The fact that $%
p\pm ip$ takes the place of $d\mp id$ is due to the flux, which breaks the
sub-valley reflection symmetry $M_x\tau_x$ and sub-valley six-fold rotation
symmetry $C_6\tau_x$. Therefore the Cooper pair of even and odd parity can
be mixed, and the orbital angular momentum of Cooper pair is discriminative
only in a modulo 3 fashion. As a result, beside $d\mp id$ pairing the $p\pm
ip$ pairing state could also occur. The trigonal distortion of the $p\pm ip$
pairing form factor can be viewed as the result of mixing with $d\mp id$.
When the flux is artificially turned off, the pairing symmetry restores the
conventional $d\mp id$. Therefore the flux is the most crucial ingredient of
new physics in this system. The nonzero orbital angular momentum of the
Cooper pair could in principle break $\mathcal{T}$. However, for the spin
triplet pairing case, it allows room for the formation of $\mathcal{T}$
related Cooper pairs.

\textit{Acknowledgment: } This work was supported by the National Key
Research and Development Program of MOST of China (No.2017YFA0302900) and by
National Natural Science Foundation of China (No. 11474331).

\end{document}